\documentclass[12pt]{article}
\usepackage{graphicx}
\def\Journal#1#2#3#4{{#1} {\bf #2}, #3 (#4)}

\def\NP{\em Nucl. Phys.}
\def\PL{\em Phys. Lett.}
\def\PRL{\em Phys. Rev. Lett.} 
\def\PR{\em Phys. Rev.}
\def\PTP{\em Prog. Theor. Phys.}
\def\ZP{\em Z. Phys.}

\begin{document}
\vskip 1.5 cm

\title{Unitarity Triangle from CP invariant quantities}

\author{
~Konrad Kleinknecht\footnote{Electronic address: 
konrad.kleinknecht@uni-mainz.de}  ~and 
Burkhard Renk\footnote{Electronic address: 
burkhard.renk@uni-mainz.de} \\
~\\
Johannes-Gutenberg Universit\"{a}t \\
Mainz \\
Germany}

\maketitle

\begin{abstract}
We construct the CKM unitarity triangle from CP invariant quantities, using the coupling constant of weak decays with flavor change from b to u, and the particle - antiparticle mixing probabilities in the ${B_s}^0 $ and ${B_d}^0 $ systems. Also included are new measurements of the coupling $V_{us}$ in Kaon decays. Of the two solutions, one agrees perfectly with the triangle constructed from CP violating processes in the K and B meson systems. The common solution yields a triangle with an area of $J/2 =  (1.51 \pm 0.09 ) \times 10^{-5}$ and a CP violating phase $ \delta = 63.1^o \pm 4.0^o$.
\end{abstract}
\newpage

The Discovery of direct CP violation in the Kaon system and the confirmation 
in the neutral B meson system showed that CP violation is due to the weak interaction \cite{KKWahl06}. In the CKM scheme 
for three generations of quarks, one phase in the complex mixing matrix can cause a difference between CP conjugate transitions. 
However, if this phase vanishes, this scheme cannot account for CP violation. The CKM mixing matrix parametrizes transitions between three generations of quarks, most of the relevant observables are derived from CP invariant 
processes, in particular weak decays of quarks.

In this letter we point out that new measurements of the mixing between ${B_s}^0 $ and ${\bar{B}_s}^0$ \cite{D0BS} \cite{CDFBS} mesons together with data on mixing between ${B_d}^0 $ and ${\bar{B}_d}^0$ and on transitions between 
b and u quarks allow a determination of the phase $\delta$ in the CKM scheme in spite of the fact that both processes are CP invariant. We then compare this double-valued result for $\delta$ with the value obtained from measurements of CP violating observables in the K meson and B meson system.

The matrix relating the quark mass eigenstates and the weak eigenstates for six quarks was introduced with an 
explicit parametrization by Kobayashi and Maskawa
{\,}\cite{KobayashiMaskawa73} in 1973. By convention, the mixing is expressed in terms of a 
$3\times 3$ unitary matrix $V$ operating on the charge $-e/3$ 
quark mass eigenstates ($d$, $s$, and $b$):
\begin{equation}
\left(\matrix{d ^{\,\prime}   \cr
                s ^{\,\prime} \cr
                b ^{\,\prime} \cr
       		}\right)
=
	\left(\matrix{
		V_{ud}&     V_{us}&     V_{ub}\cr
		V_{cd}&     V_{cs}&     V_{cb}\cr
		V_{td}&     V_{ts}&     V_{tb}\cr
       		}\right)
	\left(\matrix{
		d \cr
                s \cr
                b \cr
       		}\right) ~.
\end{equation}

We use the 
``standard'' parametrization{\,}\cite{StandardParametrization} 
of V that utilizes angles $\theta_{12}$, $\theta_{23}$, 
$\theta_{13}$, and a phase, $\delta$
\begin{equation}
V =	\left(\matrix{
	c_{_{12}} c_{_{13}}&
	s_{_{12}} c_{_{13}}&
	\: s_{_{13}} e^{-i\delta } \cr
	-s_{_{12}} c_{_{23}}
	-c_{_{12}} s_{_{23}} s_{_{13}} e^{i\delta }&
	c_{_{12}} c_{_{23}} 
        -s_{_{12}} s_{_{23}} s_{_{13}} e^{i\delta }&
	s_{_{23}} c_{_{13}}                \cr
	s_{_{12}} s_{_{23}} -c_{_{12}} c_{_{23}}
		s_{_{13}} e^{i\delta }&
	-c_{_{12}} s_{_{23}} 
	       -s_{_{12}} c_{_{23}}
		s_{_{13}} e^{i\delta } &
	c_{_{23}} c_{_{13}}                 \cr
	}\right) ~,
\label{eq:CKMparametrization} 
\end{equation}
with $c_{_{ij}} = \cos\theta_{ij}$ and $s_{ij} = \sin\theta_{ij}$ 
for the ``generation'' labels ${i,j = 1,2,3}$.  
This parametrization is exact to all orders, and has four parameters; 
the real angles $\theta_{12}$, $\theta_{23}$, $\theta_{13}$
can all be made to lie in the first quadrant
by an appropriate redefinition of quark field phases.

Information on the smallest matrix 
elements of the CKM matrix can be 
summarized in terms of the ``unitarity triangle,'' one of 
six such triangles. 
Unitarity applied to the first and third columns yields
\begin{equation}
V_{ud} ~{V_{ub}}^{\!*} + V_{cd} ~{V_{cb}}^{\!*} 
+ V_{td} ~{V_{tb}}^{\!*} = 0 ~.
\label{eq:FullUnitarityTriangle}
\end{equation}

The unitarity triangle is just a geometrical presentation 
of this equation in the complex 
plane{\,}\cite{UnitarityTriangle}.

The angles of the 
triangle $\alpha$, $\beta$ and $\gamma$  are also referred to as $\phi_2$, $\phi_1$, 
and $\phi_3$, respectively, with $\beta$ and 
$\gamma$ being the phases of the 
CKM elements $V_{td}$ and $V_{ub}$ as per 
\begin{equation}
{V_{td} = |{V_{td}}| e^{-i\beta }}, 
{V_{ub} = |{V_{ub}}| e^{-i\gamma}} ~
\end{equation}
to a precison of better than a tenth of a degree.

Rescaling the triangle so that the base is of unit length, 
the coordinates of the vertices become: 
\begin{equation}
\bigl( \bar{\rho} = \hbox{Re}(V_{ud}~V_{ub}^{\!*})/|V_{cd}~V_{cb}^{\!*}|,
~\bar{\eta} = \hbox{Im}(V_{ud}~V_{ub}^{\!*})/|V_{cd}~V_{cb}^{\!*}| \bigr),
~(1,0),
~{\rm and}~(0,0) ~.
\label{eq:ScaledUnitarityTriangle}
\end{equation}

$CP$-violating processes involve the phase $\delta$ in the 
CKM matrix, assuming that the observed $CP$ violation 
is solely related to a nonzero value of this phase.
A necessary and sufficient condition
for $CP$ violation with three generations is than that the determinant $J$ of the commutator 
of the mass matrices for the charge $2e/3$
and charge $-e/3$ quarks{\,} is non-zero \cite{Jarlskog85}.
$CP$-violating amplitudes or differences of rates are 
all proportional to the product of CKM factors in 
this quantity, namely 
$J = s_{_{12}} s_{_{13}} s_{_{23}} c_{_{12}} 
c_{_{13}}^2 c_{_{23}} \sin{\delta}$. 
This is just twice the area of the unitarity triangle.

We now proceed to determine the unitarity triangle from CP invariant quantities alone. For the following input data we refer to our last review \cite{PDGCKM04} and the references therein: $|V_{ud}| =  0.9738 \pm 0.0005$, $|V_{cd}| =  0.224 \pm 0.012$, $|V_{cs}|^2 =  (2.039 \pm 0.026) - |V_{ud}|^2 - |V_{us}|^2 - |V_{ub}|^2 - |V_{cd}|^2 - |V_{cb}|^2$, ${|V_{tb} |^2 \over |V_{td} |^2 + |V_{ts} |^2 + |V_{tb} |^2} 
= 0.94_{-0.24}^{+0.31}$.

$|V_{us} |$ --
New results from NA48 \cite{NA48KE3}, KLOE \cite{KLOEKE3}, and KTEV \cite{KTEVKE3} on the branching ratio of $K_L \rightarrow \pi e \nu$ and on the $K_L$ lifetime have been used to extract $V_{us} \times f^+(0,K^0)=0.2171 \pm 0.0004$. Two additional results from $K^+ \rightarrow \pi^0 e \nu$ yield an average $V_{us} \times f^+(0,K^+)= 0.2233 \pm 0.0011$. Two independent theoretical evaluations of the form factor $f^+(0)$ have been published. Chiral perturbation theory at the $p^6$ approximation gives $f^+(0,K^0)=0.981 \pm 0.010$ for $K^0$ and $1.002 \pm 0.010$ for $K^+$ \cite{Ciri}. Lattice calculations yield $f^+(0,K^0)=0.961 \pm 0.009$\cite{Beci}.We take the average of both and obtain $V_{us}=0.2238 \pm 0.0004 \pm 0.0023$ from $K^0$ decays and $0.2252 \pm 0.0022$ from $K^+$ decays.The overall average is
\begin{equation}
|V_{us}|  =  0.2244  \pm  0.0022 ~,
\label{eq:KdecaysVus}
\end{equation} 
where the error comes chiefly from the theoretical uncertainty.

$|V_{cb} |$ -- 
Measurements of the exclusive decay 
$B \rightarrow \bar{D}^{*}  \ell^+ \nu_{\ell}$ 
have been used to extract a value of $|V_{cb}|$ 
using corrections based on HQET \cite{IsgurWise89}.  
Exclusive $B \rightarrow \bar{D} \ell^+ \nu_{\ell}$ decays
give a consistent, but less precise result.
Analysis of inclusive decays depends on going from the quark 
to the hadron level and involves an assumption on the 
validity of quark-hadron duality. The results for $|V_{cb}|$ 
from exclusive and inclusive decays generally are in 
good agreement.  A more detailed discussion and references   
are found in a mini-review in the Review of Particle 
Physics{\,}\cite{Vcb04}.  We take the average of the 
exclusive result $|V_{cb}| =  
(41.3 \pm 1.0 \pm 1.8 ) \times 10^{-3}$ \cite{HFAG05}   and 
inclusive result $|V_{cb}| =  
(41.4 \pm 0.6 \pm 0.1 ) \times 10^{-3}$ \cite{Bauer04} with 
theoretical uncertainties combined linearly, 
weighted with their contribution to the average, and obtain

\begin{equation}
|V_{cb}| =  (41.4 \pm 0.8 ) \times 10^{-3}~.
\label{eq:Vcb06}
\end{equation}

$|V_{ub}|$ -- A compilation of the most recent results on $|V_{ub}|$ 
has been published by the "Heavy Flavour Averaging Group" recently \cite{HFAG06}, 
averaging results obtained from the CLEO, Babar and BELLE experiments. 
The quoted uncertainties for the inclusive determination is much smaller 
than for the exclusive ones, we therefore choose to use their inclusive average. 

\begin{equation}
|V_{ub}| =  (4.39 \pm 0.46 ) \times 10^{-3} ~.
\label{eq:Vub06}
\end{equation}

The new element in this analysis is the measurement of ${B_s}^0 - {\bar{B}_s}^0$ - mixing by the $ D0 $ \cite{D0BS} and $ CDF $ \cite{CDFBS} collaborations . We assume that the data shown by these experiments are evidence for such mixing. Then the mixing parameter is determined by $ D0 $ to be 

\begin{equation}
\Delta M_{B_s} = (19 \pm 1) ~{\rm ps}^{-1}
\end{equation}

This finding has been confirmed by the $CDF$ collaboration \cite{CDFBS}, they measure the mixing parameter to be
\begin{equation}
\Delta M_{B_s} = (17.33^{+0.42}_{-0.21}(stat) \pm 0.07(syst)) ~{\rm ps}^{-1}
\end{equation}

In the ratio of $B_s$ to $B_d$ mass differences, many common 
factors (such as the QCD correction and dependence on the 
top-quark mass) cancel, and we have
\begin{equation}
{\Delta M_{B_s} \over \Delta M_{B_d} } = {M_{B_s} \over M_{B_d} }
{\hat{B}_{B_s} {f_{B_s}}^2 \over \hat{B}_{B_d} {f_{B_d}}^2 }
{|{V_{tb}}^* \cdot V_{ts}|^2 \over |{V_{tb}}^* \cdot V_{td}|^2 }~.
\label{eq:BsBdMixing}
\end{equation}
With the experimentally measured masses, the well measured quantity $\Delta M_{B_d}$ \cite{Mixing04},
$\hat{B}_{B_s} {f_{B_s}}^2 /(\hat{B}_{B_d} {f_{B_d}}^2 ) 
= (1.210 ^{+0.047}_{-0.035})^2  $ {\,}\cite{Okamoto05}, these new measurements when averaged to $\Delta M_{B_s} = (17.50^{+0.40}_{-0.22}) ~{\rm ps}^{-1}$, transform into
\begin{equation}
|V_{ts}|/|V_{td}| = 4.78 \pm 0.17 ~.
\label{eq:BsMixingVtd}
\end{equation}

\begin{figure}[h]
\begin{center}
\includegraphics [width=12cm]{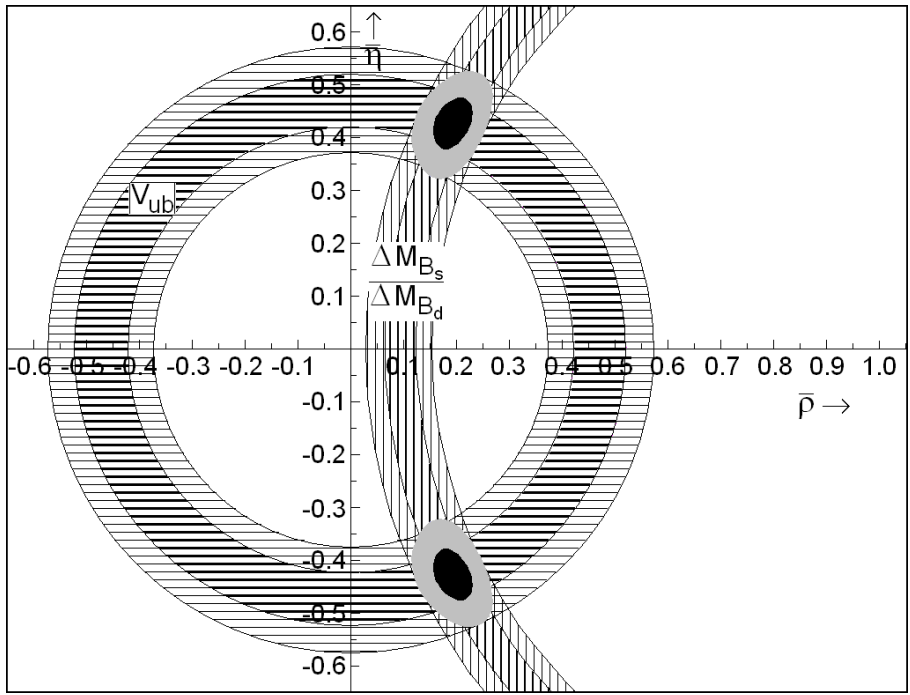}
\caption{Unitarity triangle constrained by CP conserving quantities only. Bold lines indicate one standard deviation, thin lines two standard deviations.}
\label{fig:CKM06-CPC}
\end{center}
\end{figure}

A detailed statistical analyis of all available measurements, including those from Tevatron, LEP and SLD, yields a significance of the $B_s$ mixing signal to be 3.8 $ \sigma $ \cite{HFAGweb}.  We insert equation \ref{eq:BsMixingVtd} in our fit of the CKM matrix and obtain contours for the allowed range of the top of the unitarity triangle, at one and two standard deviation confidence level, as shown in Figure \ref{fig:CKM06-CPC} . The $\chi^2_{min}$ found is 3.3 with 4 degrees of freedom, this or a higher value is expected to appear with 50\% probability.  We define a one standard deviation confidence level to be the multitude of all points in the unitarity plane to have a solution in the four dimensional parameter space with a $\chi^2 < \chi^2_{min} + 1$, and the two standard deviation confidence to fulfil $\chi^2 < \chi^2_{min} + 4$ respectively. This is a slightly different approach than the one used by other groups \cite{CKMfitter}, \cite{UTfit}, leading to similar results given the precision at which the larger angles are known.

Both measurements of $|V_{ub}|$ and $|V_{ts}|/|V_{td}|$ yield
a circle-shaped range of allowed values. The two circles overlap
at two positions corresponding to $ \delta = 66^o \pm 4^o$ 
and $ \delta = 294^o \pm 4^o$ . The measurements of CP invariant 
quantities alone therefore allow for the first time the prediction 
that CP violation exists in the framework of the CKM scheme. 
If the two circles would overlap at $ \delta = 0 $ or
$ \delta = 180^o $, no CP violation would be expected. 
This would correspond to values of 
$|V_{ts}|/|V_{td}| = 7.7$ resp. $|V_{ts}|/|V_{td}| = 3.1$ 
Only the latter of these solutions was excluded by earlier 
measurements, the first was still allowed within two standard 
deviations from the measurement of $\Delta M_{B_d}$ and 
lattice QCD calculations.

\begin{figure}[h]
\begin{center}
\includegraphics [width=12cm]{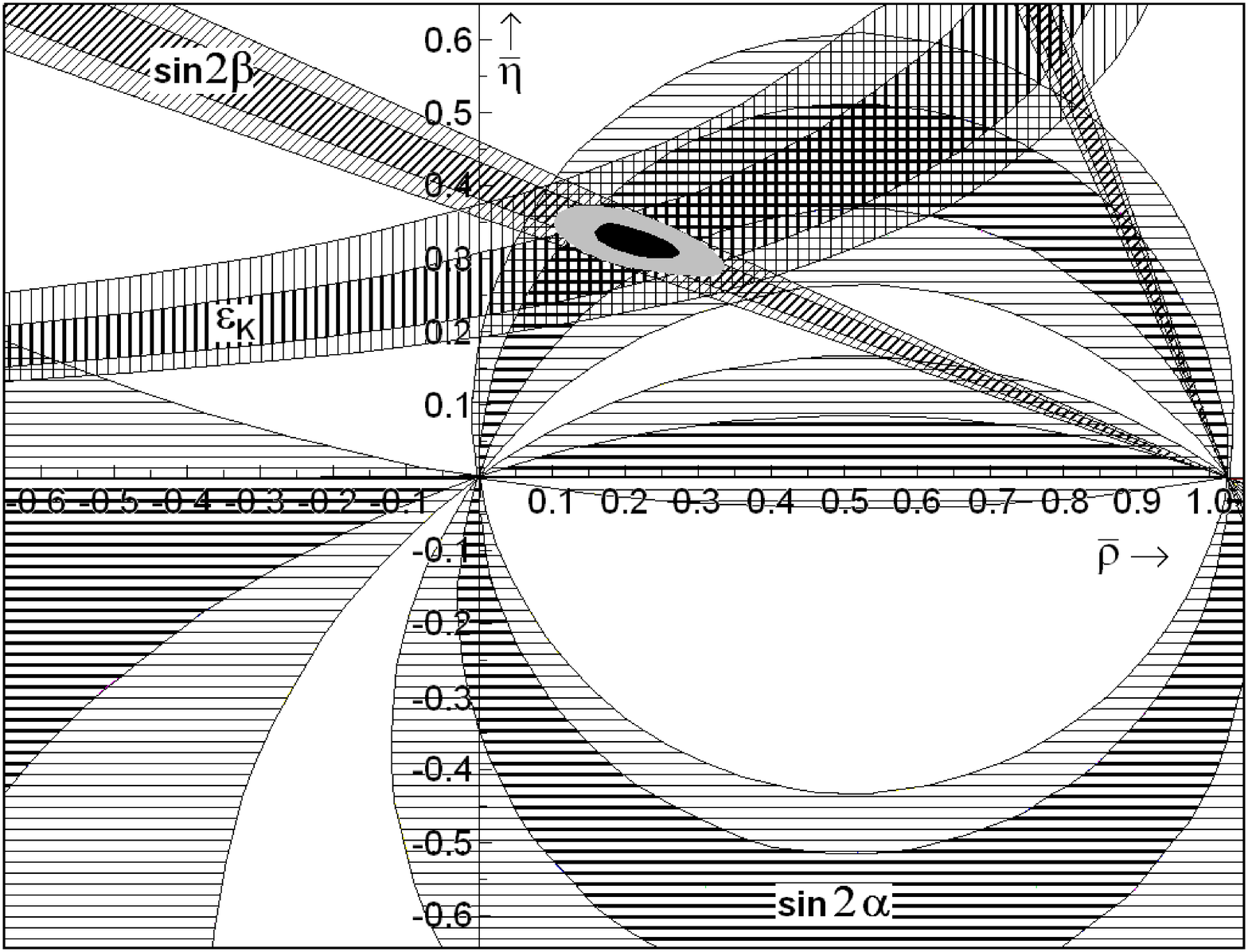}
\caption{Unitarity triangle constrained by CP violating quantities only. Bold lines indicate one standard deviation, thin lines two standard deviations.}
\label{fig:CKM06-CPV}
\end{center}
\end{figure}

We now turn to CP violating effects. Three well understood independent 
measurements are available by now, the direct measurements
of $\gamma$ in $ B \rightarrow  D^{(+)} K$ still have too
large uncertainties . Just the added constraint from $CP$ violation in the neutral 
kaon system, taken together with the restrictions above 
on the magnitudes of the CKM matrix elements, is tight 
enough to restrict considerably
the range of angles and the phase of the CKM matrix.  
For example, the constraint obtained from the $CP$-violating 
parameter $\epsilon_K$ in the neutral~$K$ system corresponds to 
the vertex~A of the unitarity triangle lying on a hyperbola 
for fixed values of the (imprecisely known) hadronic matrix 
elements{\,}\cite{epsilonKQCD}{,\,}\cite{BKparameter}.

In the B-meson system, for $CP$-violating asymmetries 
of neutral $B$ mesons decaying to $CP$ eigenstates, 
the interference between mixing and a single weak 
decay amplitude for certain final states directly relates 
the asymmetry in a given decay to $\sin 2\phi$, 
where $\phi = \alpha$, $\beta$, $\gamma$ is an appropriate 
angle of the unitarity triangle{\,}\cite{UnitarityTriangle}. 
A new generation of experiments has established a 
non-vanishing asymmetry in the decays 
$B_d (\bar{B}_d ) \rightarrow \psi K_S$ and in other $B_d$ 
decay modes where the asymmetry is given by $\sin 2\beta$.  
The present experimental results from 
BaBar{\,}\cite{BabarSine2beta05} and 
Belle{\,}\cite{BelleSine2beta05}, when averaged yield
\begin{equation}
\sin 2\beta = 0.687 \pm 0.032 ~.
\label{eq:sine2beta}
\end{equation}

A non vanishing asymmetry in the decays $B_d (\bar{B}_d ) \rightarrow \ \rho \rho $ from 
BaBar{\,}\cite{BabarSine2alpha05} and 
Belle{\,}\cite{BelleSine2alpha05}, $B_d (\bar{B}_d ) \rightarrow \ \rho \pi $ and $B_d (\bar{B}_d ) \rightarrow \ \pi \pi $,
when averaged, can be used to constrain the angle $\alpha$.

The confidence level contours from the $\pi\pi$ and $\rho\rho$ isospin analyses as well 
as the $\rho\pi$ Dalitz plot yield \cite{HFAGweb} 

\begin{equation}
\alpha = (99 ^{+12} _{-9} [1\sigma] ^{+22} _{-16} [2\sigma])^o ~.
\label{eq:sine2alpha}
\end{equation}

\begin{figure}[h]
\begin{center}
\includegraphics [width=12cm]{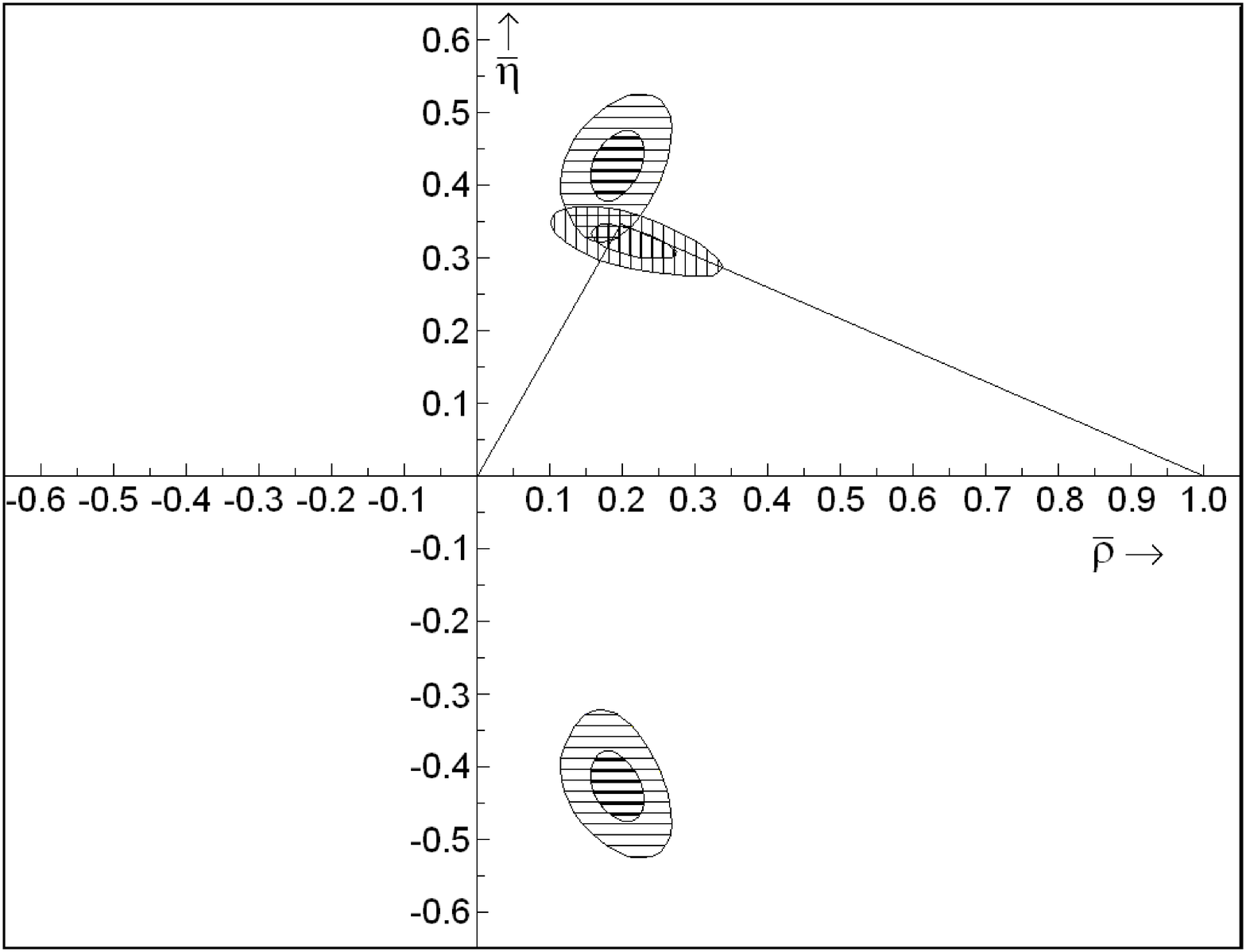}
\caption{Comparison of unitarity triangle apex obtained from CP concerving amplitudes (horizontally dashed )to the one obtained from CP violating quantities (vertically dashed). Also shown is the best fit unitarity triangle.}
\label{fig:CPCvsCPV}
\end{center}
\end{figure}

These three constraints, together with those on the large angles, applied to the unitarity plane, overlap only in one region at the one or two standard deviation level, as shown in figure  
\ref{fig:CKM06-CPV}. The resulting CP-phase $ \gamma = 56.7^o \pm 8.0^o$, as measured from CP violating effects, can be compared with the one predicted from the CP conserving measurements, as shown in Figure  \ref{fig:CPCvsCPV}. 

An overall fit of the CKM matrix yields 
values of the sines of the angles of 
$s_{_{12}} = 0.2261 \pm 0.0014 $, 
$s_{_{23}} = 0.04125 \pm 0.00070 $, 
and $s_{_{13}} = 0.00362 \pm 0.00018$ and to a value of the CP-phase $ \gamma = 63.1^o \pm 4.0^o$, with $\chi^2_{min} = 7.1$ for 7 degrees of freedom, this or a higher value expected at the 42\% level. Finally, $V_{td }= (0.00774 \pm 0.00027 ) - (0.00314 \pm 0.00018 ) i $. The area of the unitarity triangle is $J/2 =  (1.51 \pm 0.09 ) \times 10^{-5}$. 

The two allowed regions in the upper half-plane of Figure  \ref{fig:CPCvsCPV} overlap at better than two standard deviations. The combined fit has 7 degrees of freedom, 3 more than the fit to CP conserving quantities alone. The $\chi^2$ value increases from 3.3 to 7.1 by adding the constraints from CP violating quanties. In conclusion, we have shown that after the observation of $B_s$ mixing, the unitarity triangle can be constructed from CP conserving quantities alone, and one of the resulting solutions agrees at the 50\% confidence level with the triangle constructed from CP conserving and CP violating quantities. 

~\\
\noindent{\bf ACKNOWLEDGMENT} \\
~\\
This research work is supported in part by the Bundesministerium f\"ur 
Bildung und Forschung under Grant No. 05HE1UMA/3. We thank H. Wittig for a critical discussion of lattice results.


\begin{thebibliography} {99}

\bibitem{KKWahl06} K. Kleinknecht and H.Wahl, Eur.Phys.News vol.37, p.26-29 (2006); K.Kleinknecht, "Uncovering CP Violation, Experimental clarification in the Neutral K Meson and B Meson systems", Springer tracts in Modern physics, vol.195 (2003)

\bibitem{D0BS} V.M.Abazov {\em et al.},hep-ex/0603029, 2006.

\bibitem{CDFBS} G.Gomez-Ceballos {\em et al.}, FPCP2006 conference, Vancouver, April 2006; http://www-cdf.fnal.gov/physics/new/bottom/060406.blessed-Bsmix/ .

\bibitem{KobayashiMaskawa73}M. Kobayashi and T. Maskawa,
	\Journal{\PTP} {49} {652} {1973}.

\bibitem{StandardParametrization}L.-L. Chau and W.-Y. Keung,
	\Journal{\PRL} {53} {1802} {1984};
	H. Harari and M Leurer, 
	\Journal{\PL} {B181} {123} {1986};
	H. Fritzsch and J. Plankl, 
	\Journal{\PR} {D35} {1732} {1987};
	F. J. Botella and L.-L. Chao, 
	\Journal{\PL} {B168} {97} {1986}.

\bibitem{Wolfenstein83}L. Wolfenstein, 
	\Journal{\PRL} {51} {1945} {1983}.

\bibitem{Jarlskog85}C. Jarlskog, 
	\Journal{\PRL} {55} {1039} {1985} and 
	\Journal{\ZP} {C29} {491} {1985}.

\bibitem{UnitarityTriangle}L.-L. Chau and W. Y. Keung, Ref.~3; 
	J. D. Bjorken, private communication and
	\Journal{\PR} {D39} {1396} {1989};
	C. Jarlskog and R. Stora, 
	\Journal{\PL} {B208} {268} {1988};
	J. L. Rosner, A. I. Sanda, and M. P. Schmidt, in
	{\em Proceedings of the Workshop on High Sensitivity
	Beauty Physics at Fermilab}, Fermilab, November 11 - 14,
	1987, edited by A. J. Slaughter, N. Lockyer, and
	M. Schmidt (Fermilab, Batavia, 1988), p. 165;
	C. Hamzaoui, J. L. Rosner, and A. I. Sanda, 
	{\em ibid.}, p. 215.    

\bibitem{PDGCKM04} F. Gilman, K. Kleinknecht and B. Renk, in S. Eidelman et al., \Journal{\PL} {B592} {1} {2004} and 2005 partial update for the 2006 edition available on the PDG WWW pages (URL: http://pdg.lbl.gov/)

\bibitem{NA48KE3} A. Lai {\em et al.}, \Journal{\PL} {B602} {41} {2004}.

\bibitem{KLOEKE3} F.Ambrosino {\em et al.},PoS HEP2005:287, 2006

\bibitem{KTEVKE3} T. Alexopoulos {\em et al.}, \Journal{\PRL} {93} {181802} {2004}.

\bibitem{Ciri} V. Cirigliano, H. Neufeld and H. Pichl, Eur.\ Phys.\
  J.\ C 35, 53 (2004) 
  
\bibitem{Beci} D. Becirevic et al., hep-ph/04003217


\bibitem{IsgurWise89}N. Isgur and M. B. Wise, 
	\Journal{\PL} {B232} {113} {1989} and 
	\Journal{\PL} {B237} {527} {1990} E;  
	E. Eichten and B. Hill,
	\Journal{\PL} {B234} {511} {1990};
	M. E. Luke, \Journal{\PL} {B252} {447} {1990}.

\bibitem{Vcb04} M. Artuso and E. Barberio, mini-review of 
	$V_{cb}$ in the 2004 Review of Particle Physics.


\bibitem{HFAG05}K. Anikeev {\em et al.} 
 	(Heavy Flavour Averaging group, HFAG ), hep-ex/0505100, 2005.
 	
\bibitem{CKMfitter}CKMfitter Group (J. Charles et al.), 
Eur. Phys. J. C41, 1-131 (2005), and http://www.slac.stanford.edu/xorg/ckmfitter.

\bibitem{UTfit}UTfit Collaboration (M. Bona et al.),
JHEP 0603 (2006) 080, and www.utfit.org.


 	
\bibitem{Bauer04}C. W. Bauer {\em et al.} 
 	\Journal{\PR} {D70} {094017} 2004.

\bibitem{HFAG06}E. Barberio {\em et al.} 
 	(Heavy Flavour Averaging group, HFAG ), hep-ex/0603003, 2006.


\bibitem{Fitsbyus}K. Kleinknecht and B. Renk, 
	\Journal{\PL} {86} {130B} {1983}
	\Journal{\ZP} {C34} {209} {1987}.

\bibitem{Mixing04}O. Schneider, mini-review on $B$ - $\bar{B}$ 
 	mixing in the 2004 Review of Particle Physics.
 	
\bibitem{HFAGweb}HFAG, unpublished, see http://www.slac.stanford.edu/xorg/hfag/triangle/index.html .

\bibitem{Okamoto05} M. Okamoto, PoS (Lat2005)913, XXIIIrd International Symposium on Lattice Field Theory, Dublin, Ireland , hep-lat/0510113.

\bibitem{Buras90}A. J. Buras {\em et al.}, 
	\Journal{\NP} {B347} {491} {1990}.

\bibitem{bsgammacleo}M. S. Alam  {\em et al.} (CLEO Collaboration),
	\Journal{\PRL} {74} {2885} {1995}; 
      S. Chen {\em et al.} (CLEO Collaboration),
	\Journal{\PRL} {87} {1807} {2001}

\bibitem{bsgammabelle}T. Abe  {\em et al.} (Belle Collaboration),
	\Journal{\PL} {B511} {157} {2001}

\bibitem{Ali03}A. Ali and M. Misiak, 
	hep-ph/0304132, 2003.

\bibitem{KplusEvents}S. Adler {\em et al.} (E787 Collaboration), 
	hep-ex/0111091, 2001.

\bibitem{epsilonKQCD}The relevant QCD corrections 
	in leading order in F. J. Gilman and M. B. Wise
	\Journal{\PL} {B93} {129} {1980} and
	\Journal{\PR} {D27} {1128} {1983}, have been
	extended to next-to-leading-order by
	A. Buras {\em et al.}, Ref. 39;  
	S. Herrlich and U. Nierste
	\Journal{\NP} {B419} {292} {1992} and 
	\Journal{\NP} {B476} {27} {1996}.

\bibitem{BKparameter}The limiting curves in Figure 2
	arising from the value of $|\epsilon |$ correspond
	to values of the hadronic matrix element expressed 
      in terms of the renormalization group invariant 
	parameter $\hat{B}_K$ from 0.74 to 0.98 for the 1 SD contour, and 0.62 to 1.1 for the 2 SD contour.~.  See, 
	for example, D. Becirevic, plenary talk at Lattice 2003, 
	Tsukuba, Japan, July 15 - 19, 2003.


\bibitem{epsilonprimenonzero}H. Burkhardt {\em et al.}, 
	\Journal {\PL} {B206} {169} {1988};
 	A. Alavi-Harati {\em et al.}, 
	\Journal{\PR} {D67} {012005} {2003};
	J.R. Batley {\em et al.}, \Journal{\PL }{B544} {97} {2002}; 
	H.Wahl, Phys.Rept.403-404 19-25 {2004} 
.

\bibitem{BabarSine2beta05}B. Aubert {\em et al.},  	
	\Journal{\PRL} {94} {161803} {2005}.

\bibitem{BelleSine2beta05}K. Abe {\em et al.} 
	(Belle Collaboration), Belle -CONF-0569, hep-ex/0507037. 

\bibitem{BabarSine2alpha05}B. Aubert {\em et al.},  	
	\Journal{\PRL} {95} {041805} {2005}.

\bibitem{BelleSine2alpha05}K. Abe {\em et al.} 
	(Belle Collaboration), Belle -CONF-0545, hep-ex/0507039. 

\end{thebibliography}
\end{document}